\documentclass[fleqn,11pt]{article}

\usepackage{epsf}
\usepackage{amsmath,amssymb}

\usepackage{amsfonts,amssymb,cite}
\usepackage{epsf,graphicx}

\newcommand{\bls}[1]{\renewcommand{\baselinestretch}{#1}}

\def\noi{\noindent}
\def\jnumber#1#2{\thispagestyle{empty} \noi\unitlength=1mm
    	\begin{picture}(178,10)
            \put(177,15){\llap{\large\it Grav. Cosmol. No.\,#1, #2}}
                    \end{picture}}

\newcommand{\Title}[1]{\noi {{\Large\bf #1}}\\[1ex]}

\newcommand{\Author}[2]{\noi{\bf #1}\\[2ex]\noi{\normalsize\it #2}\\}

\newcommand{\Rec}[1]{\noi {\it Received #1} \\}

\newcommand{\Abstract}[1]{\vskip 2mm \begin{center}
        \parbox{16.4cm}{\small\noi #1} \end{center}\medskip}

\def\nqq{\hspace*{-2em}}

\def\Jl#1#2{#1 {\bf #2},\ }

\def\ApJ#1 {\Jl{Astroph. J.}{#1}}
\def\CQG#1 {\Jl{Class. Quantum Grav.}{#1}}
\def\DAN#1 {\Jl{Dokl. AN SSSR}{#1}}
\def\GC#1 {\Jl{Grav. Cosmol.}{#1}}
\def\GRG#1 {\Jl{Gen. Rel. Grav.}{#1}}
\def\JETF#1 {\Jl{Zh. Eksp. Teor. Fiz.}{#1}}
\def\JETP#1 {\Jl{Sov. Phys. JETP}{#1}}
\def\JHEP#1 {\Jl{JHEP}{#1}}
\def\JMP#1 {\Jl{J. Math. Phys.}{#1}}
\def\NPB#1 {\Jl{Nucl. Phys. B}{#1}}
\def\NP#1 {\Jl{Nucl. Phys.}{#1}}
\def\PLA#1 {\Jl{Phys. Lett. A}{#1}}
\def\PLB#1 {\Jl{Phys. Lett. B}{#1}}
\def\PRD#1 {\Jl{Phys. Rev. D}{#1}}
\def\PRL#1 {\Jl{Phys. Rev. Lett.}{#1}}

\def\lal{&&\nqq {}}

\def\beq{\begin{equation}}
\def\eeq{\end{equation}}
\def\bear{\begin{eqnarray}}
\def\bearr{\begin{eqnarray} \lal}
\def\ear{\end{eqnarray}}
\def\earn{\nonumber \end{eqnarray}}

\renewcommand{\lim}{\mathop{\rm lim}\limits}

\bls{0.98}
\begin{document}
\twocolumn[
\jnumber{XX}{2014}

\vspace{-6mm}

\Title{Geometric Origin of Pair Production by Electric Field in de Sitter Space}

\Author{Sang Pyo Kim}
{Department of Physics, Kunsan National University, Kunsan 573-701, Korea}

\Rec{April 15, 2014}

\Abstract
   {The particle production in a de Sitter space provides an interesting model to understand the curvature effect on Schwinger pair production by a constant electric field or Schwinger mechanism on the de Sitter radiation. For that purpose, we employ the recently introduced complex analysis method, in which the quantum evolution in the complex time explains the pair production via the geometric transition amplitude and gives the pair-production rate as the contour integral. We compare the result by the contour integral with that of the phase-integral method.}

] 

{
\def\os{\overline{s}}
\def\oC{\overline{C}}
\def\cL{\mathcal{L}}
\def\cS{\mathcal{S}}

\section{Introduction} \label{sec 1}

A background electromagnetic field probes the vacuum structure of Dirac sea of virtual charged pairs and a spacetime curvature also changes the vacuum structure. Schwinger mechanism of pair production and the vacuum polarization in the elecromagnetic field is the most prominent feature of the nonperturbative quantum effects \cite{Schwinger51}. In fact, the one-loop effective action in the electromagentic field with a nonvanishing electric field is complex, whose imaginary part is related to the pair production \cite{DeWitt75,DeWitt03}. In a curved spacetime particles are produced due to the presence of a horizon or nonstationary nature of the spacetime itself, of which Hawking radiation from a black hole and Gibbons-Hawking radiation from a de Sitter space are the most well-known phenomenon.

Recently the author has introduced a complex analysis method for particle production in an electric field or a de Sitter space by expressing the production rate as the contour integral of the frequency in the complex plane of time \cite{Kim13a,Kim13b,Kim14a}. The particle-production rate is given by \cite{Kim-Bull}
\begin{eqnarray}
{\cal N}_{\kappa} = \Big\vert \sum_{J} \exp \Bigl[- i \oint_{C_{J}^{(1)} (t_0)} \omega_{\kappa} (z) dz \Bigr] \Big\vert,
\label{pp rate}
\end{eqnarray}
where $\omega_{\kappa} (z)$ is the frequency with the quantum number $\kappa$ of a quantum field in the complex plane, $C_{J}^{(1)} (t_0)$ exhausts all the classes of closed loops of winding number one with the base point $t_0$. The particle-production rate (\ref{pp rate}) is one method to compute the Bogoliubov coefficient. The conventional wisdom has introduced the in-vacuum and the out-vacuum and found the Bogoliubov coefficients between two vacua. The complex analysis has also been used in the phase-integral method, which gives the probability for scattering over and under the potential barrier \cite{Kim-Page02,Kim-Page07,Dumlu-Dunne10,Dumlu-Dunne11a,Dumlu-Dunne11b}.

In this paper, we critically review the complex analysis method for particle production and then apply the formula (\ref{pp rate}) to pair production of charged scalars by a constant electric field in the (1+1)-dimensional de Sitter space. By doing so, we may see how Schwinger mechanism and Gibbons-Hawking mechanism are intertwined for the radiation. Schwinger pair production has been studied in the de Sitter space \cite{Garriga94,Villaba95,Kim-Page08,KHW11,Haouat-Chekireb13,KYHS14}.
What is new in this paper is that we study the pair production using the formula (\ref{pp rate}) and clarify the origin of particle production. Finally, we compare the contour integral method with the phase-integral method.

\section{Charged Scalars in Electric Field in de Sitter Space}  \label{sec 2}

We consider a charged scalar field in a constant electric field in the (1+1)-dimensional planar coordinates of de Sitter space [in units of $\hbar = c =1$]
\begin{eqnarray}
ds^2 = - dt^2 + a^2(t) d{\bf x}^2, \quad a(t) = e^{Ht}. \label{ds}
\end{eqnarray}
The constant electric field has the vector potential in a one-form \cite{Kim-Page08}
\begin{eqnarray}
A (t) = - \frac{E}{H} \bigl(e^{Ht} - 1 \bigr) dx,
\end{eqnarray}
where a constant is subtracted to have the Minkowski limit when $H = 0$,
and the field tensor in a two-form \cite{Kim-Page08}
\begin{eqnarray}
{\bf F} = dA = E \sqrt{-g} dx \wedge dt.
\end{eqnarray}
Then the charged scalar field has the Hamiltonian \cite{Kim14b}
\begin{eqnarray}
H = \int dx \Bigl[\frac{1}{a} |\pi|^2 + \frac{1}{a} |
(\partial_x - i q A) \phi|^2 + m^2 a | \phi |^2 \Bigr], \label{action}
\end{eqnarray}
where the conjugate momenta are $\pi = a \dot{\phi}^*$ and $\pi^* = a \dot{\phi}$.
Note that Eq. (\ref{action}) is the standard QED action with the minimal interaction in a curved spacetime but other theory describing the interaction between gravity and electromagnetism, for instance Ref. \cite{Romashka}, may be used.
In terms of the Fourier modes
\begin{eqnarray}
\phi (t,x) = \int \frac{dk}{2\pi} \phi_k e^{ikx}, \quad \phi^* (t,x) = \int \frac{dk}{2\pi} \phi^*_k e^{ikx}
\end{eqnarray}
and similarly for $\pi (t,x)$ and $\pi^* (t,x)$,
the Hamiltonian is the sum of time-dependent oscillators
\begin{eqnarray}
H (t) := \sum_{k} H_k (t) = \int \frac{dk}{2\pi} \Bigl[\frac{1}{a} \pi^*_{k} \pi_k + a \omega_k^2 \phi^*_k \phi_k \Bigr], \label{ham}
\end{eqnarray}
where
\begin{eqnarray}
\omega_k^2 = \frac{1}{a^2} \Bigl(k - \frac{qE}{H}+ \frac{q E}{H} a \Bigr)^2 + m^2. \label{freq}
\end{eqnarray}

\section{Real-Time Evolution} \label{sec 3}

The quantum theory in the real-time formalism is the time-dependent (functional) Schr\"{o}dinger equation
\begin{eqnarray}
i \frac{\partial}{\partial t} \vert \Psi (t) \rangle = \hat{H}(t) \vert \Psi (t) \rangle, \label{sch eq}
\end{eqnarray}
where the quantum state of the field is the product of the quantum state for each oscillator $\hat{H}_k (t)$
\begin{eqnarray}
\vert \Psi (t) \rangle = \prod_{k} \vert \psi_k (t) \rangle.
\end{eqnarray}
To find the quantum states, we may use the quantum invariants known as Lewis-Riesenfeld invariants \cite{Lewis-Riesenfeld69}
\begin{eqnarray}
i \frac{\partial \hat{I}_k (t)}{\partial t} + [\hat{I}_k (t), \hat{H}_k (t)] = 0.
\end{eqnarray}
 A pair of quantum invariants, which play the role of the time-dependent annihilation and creation operators, constructs the time-dependent vacuum states and the excited number states \cite{MMT70,Kim-Page01}.

Following Ref. \cite{Kim14b}, we may introduce the linear quantum invariant operators for particles and antiparticles, respectively,
\begin{eqnarray}
\hat{c}_k (t) &=& i \bigl(\varphi^*_k (t) \hat{\pi}^*_k - \dot{\varphi}^*_k (t) \hat{\phi}_k  \bigr), \nonumber\\
\hat{d}_k (t) &=& i \bigl(\varphi^*_k (t) \hat{\pi}_k - \dot{\varphi}^*_k (t) \hat{\phi}^*_k  \bigr),  \label{pa inv}
\end{eqnarray}
and their Hermitian conjugates, where the auxiliary field satisfies the field equation
\begin{eqnarray}
\ddot{\varphi}_k + \omega_k^2 (t) \varphi_k = 0. \label{mod eq}
\end{eqnarray}
The operators (\ref{pa inv}) become the time-dependent annihilation operators for particles and anti-particles when the Wronskian condition from the quantization rule holds
\begin{eqnarray}
a(t) {\rm Wr} [\varphi_k (t), \varphi^*_k (t) ] = i. \label{wr}
\end{eqnarray}
The time-dependent vacuum state is annihilated by the annihilation operators for all the momenta
\begin{eqnarray}
\hat{c}_k (t) \vert 0, t \rangle = \hat{d}_k (t) \vert 0, t \rangle = 0.
\end{eqnarray}
In the in-out formalism, the scattering matrix between the in-vacuum in the asymptotic past and the out-vacuum in the asymptotic future gives the vacuum persistence amplitude \cite{DeWitt03}
\begin{eqnarray}
\langle 0, t = \infty \vert 0, t = - \infty \rangle = e^{i \int dt dx {\cal L}_{\rm eff}},  \end{eqnarray}
and the imaginary part of the effective action relates the pair production via the Bogoliubov transformation
\begin{eqnarray}
e^{ - 2 \int dt dx {\rm Im} {\cal L}_{\rm eff}} = e^{- T \int \frac{dk}{2 \pi} \ln (1+ {\cal N}_k)}.
\end{eqnarray}
The positive and negative frequency solutions satisfying Eq. (\ref{wr}) are given by \cite{Whittaker-Watson63}
\begin{eqnarray}
\varphi^{(+)}_k (t) &=& \frac{e^{ \pi \lambda/2}}{\sqrt{2\bar{k}}} W_{ - i \lambda, i \mu} (- 2i \frac{\bar{k}}{H} e^{-Ht}), \nonumber\\
\varphi^{(-)}_k (t) &=& \frac{e^{ \pi \lambda/2}}{\sqrt{2\bar{k}}} W_{ i \lambda, i \mu} ( 2i \frac{\bar{k}}{H} e^{-Ht}), \label{pos-neg}
\end{eqnarray}
where $W$ is the Whittaker function and
\begin{eqnarray}
\lambda = \frac{qE}{H}, \quad \mu = \sqrt{\Bigl( \frac{qE}{H} \Bigr)^2 + m^2 - \frac{1}{4}},
\end{eqnarray}
and $\bar{k} = k - (qE)/H$ is the shifted momentum. The solution differs from Eq. (19) of Ref. \cite{Garriga94} by the normalization constant.

The evolution operator for each oscillator is expressed in the time-ordered integral
\begin{eqnarray}
\vert \psi_k (t) \rangle = {\rm T} \exp \Bigl[ - i \int_{t_0}^{t} dt' \hat{H}_k (t') \Bigr] \vert \psi_k (t_0) \rangle. \label{ev op}
\end{eqnarray}
To have the matrix representation of the evolution operator \cite{Kim13c}, we may use
the basis of instantaneous number states
\begin{eqnarray}
\hat{H}_k (t) \vert n_k, t \rangle = \omega_k (t) \Bigl(n_k + \frac{1}{2} \Bigr) \vert n_k, t \rangle.
\end{eqnarray}
In that representation the Hamiltonian has the diagonal matrix 
\begin{eqnarray}
{\bf H}_{k} = \omega_k \bigl(\frac{1}{2}, \frac{3}{2}, \cdots \frac{2n+1}{2}, \cdots \bigr),
\end{eqnarray}
and the basis induces the gauge potential
\begin{eqnarray}
({\bf A}_k)_{m_k n_k} = i \frac{\dot{\omega}_k}{4 \omega_k} \bigl(\sqrt{n_k(n_k-1)}\delta_{m_k n_k-2} \nonumber\\ - \sqrt{(n_k+1)(n_k+2)} \delta_{m_k n_k+2} \bigr).
\end{eqnarray} 
Therefore, there is no contribution to the vacuum persistence
if there are not either simple poles or zeros of $\omega_k$ along the real-time axis. Otherwise,
the evolution of the vacuum state along a path $C$ on the real-time axis gives a trivial result $\langle 0_k, C(t = - \infty) \vert 0_k, t = - \infty \rangle = 1$.
Hence, in the in-in formalism the particle production could be understood by extending the quantum evolution to the complex plane of time as will be shown in the next section.

\section{Quantum Evolution in the Complex Plane} \label{sec 4}

To extend the quantum evolution (\ref{sch eq}) to the complex plane $z$, we assume that \cite{Kim14a}
\begin{itemize}
  \item  there exist an analytical operator $\hat{H}_k (z)$ and an analytical function $\omega_k (z)$  in the complex plane that reduce to $\hat{H}_k (t)$ and $\omega_k (t)$, respectively, on the real-time axis,
  \item  $\hat{H}_k (z)$ has an orthonormal basis $\{ \vert n_k, z \rangle \}$ with $\langle n_k, z \vert m_k, z \rangle = \delta_{n_k m_k} $ that reduces to  $\{ \vert n_k, t \rangle \}$ on the real-time axis,
  \item there exists a conformal mapping from the complex $t$ to the whole complex plane $z$.
\end{itemize}
The above conditions can be satisfied for the Hamiltonian (\ref{ham}) in the de Sitter space since $a (t) = e^{Ht}$ has an analytical continuation in the complex plane of time. Furthermore, the othornormal basis can be constructed in the complex plane from those operators (\ref{pa inv}) with $\varphi^{(+)}_k (t)$ for $\varphi_k (t)$ and $\varphi^{(-)}_k (t)$ for $\varphi^*_k (t)$ since
the positive and negative frequency solutions satisfy the Wronskian condition in the complex plane \cite{Whittaker-Watson63}.

Thus, the scattering matrix between the in-vacuum at $t = - \infty$ and another in-vacuum transported along a closed clockwise path $C_{J}(z)$ and returned to the same base point $t = - \infty$ in the complex plane is to the lowest order of the Magnus expansion given by
\begin{eqnarray}
\langle 0_k, t_0 \vert 0_k, C_{J} (- \infty) \rangle = \exp \Bigl[- \frac{i}{2} \oint_{C_{J} (-\infty)} dz \omega_k (z) \Bigr]. \label{sc am}
\end{eqnarray}
Then the production rate for one pair (\ref{pp rate}) is the sum of the scattering matrix squared over all independent paths $C^{(1)}_{J}$ which are classified by the homotopy class for the simple poles and have the winding number 1. The higher winding number corresponds to multiple pair production.

\section{Geometric Transition for Pair Production}  \label{sec 3}

The Hamiltonian (\ref{ham}) and the frequency (\ref{freq}) in the de Sitter space
have an analytical continuation by introducing the conformal mapping
\begin{eqnarray}
z = e^{Ht},
\end{eqnarray}
which covers the whole complex plane for a strip $t = [0, 2\pi i/H]$ along the real axis.
Then, the frequency is an analytical function
\begin{eqnarray}
\omega_k (z) = \sqrt{ \frac{1}{z^2}\Bigl(\bar{k} + \frac{qE}{H} z\Bigr)^{2} + m^2}, \label{an freq}
\end{eqnarray}
and the contour integral is given by
\begin{eqnarray}
\frac{1}{H} \oint \frac{dz}{z} \omega_k (z)
\end{eqnarray}
In the large $z$-expansion limit, the contour integral takes the form
\begin{eqnarray}
\frac{\bar{\mu}}{H} \oint \frac{dz}{z} \Bigl[ 1 + \frac{\bar{k}qE}{H \bar{\mu}} \frac{1}{z} + {\cal O}
\Bigl(\frac{1}{z^2} \Bigr) \Bigr] = - 2 \pi i \frac{\bar{\mu}}{H},
\end{eqnarray}
where the minus sign is due to the simple pole at infinity and
\begin{eqnarray}
\bar{\mu} = \sqrt{\Bigl( \frac{qE}{H} \Bigr)^2 + m^2}.
\end{eqnarray}
In the small $z$-expansion limit, the contour integral leads to
\begin{eqnarray}
\frac{1}{H} \oint \frac{dz}{z} \Bigl[\frac{\bar{k}}{z}  + \frac{qE}{H} + \frac{m^2}{2} \frac{z}{ \bar{k}} + {\cal O}
\Bigl(z^2 \Bigr) \Bigr] = 2 \pi i \frac{qE}{H^2}.
\end{eqnarray}
Therefore, the pair-production rate (\ref{pp rate}) by a constant electric field in the de Sitter space is given by
\begin{eqnarray}
{\cal N}_k =   \exp \Bigl[ - \frac{2 \pi}{H} \Bigl( \bar{\mu} - \frac{qE}{H} \Bigr) \Bigr]. \label{pp ds}
\end{eqnarray}
The pair-production rate (\ref{pp ds}) is consistent with the leading term of Refs. \cite{Garriga94,Kim-Page08,KHW11}.

We now compare the contour integral method with the phase-integral method, which reads \cite{Kim-Page07,Kim10}
\begin{eqnarray}
{\cal N}_k = \exp \Bigl[ - 2 {\rm Im} \int_{z^*_{k(1)}}^{z^*_{k(2)}} \omega_k (z) dz \Bigr],
\end{eqnarray}
where $z^*_{k}$ denotes a turning point in the complex plane, which is a root $\omega_k (z^*_{k}) = 0$,
\begin{eqnarray}
z^*_k = - \frac{\bar{k} qE}{H \bar{\mu}^2} \pm i \frac{\bar{k} m}{\bar{\mu}^2}.
\end{eqnarray}
A direct integration by quadrature gives
\begin{eqnarray}
\int_{z^*_{k(1)}}^{z^*_{k(2)}} dz \omega_k (z) = i \pi  \Bigl( \frac{\bar{\mu}}{H} - \frac{qE}{H^2} \Bigr).
\end{eqnarray}
The phase-integral method gives the same result as the contour integral method. However, the advantage of the complex analysis method is the computational usefulness and diversity of contour integrals, which can be done for many models for which the phase-integral may not be explicitly performed.

\section{Conclusion}  \label{sec 5}

In this paper we have applied the complex analysis method to pair production by a constant electric field in the (1+1)-dimensional de Sitter space. In the in-in formalism, the real-time dynamics gives a null result while the complex-time evolution explains the particle production as originating from the pole structure of the frequency. We have shown that the pair-production rate expressed by the contour integral gives the correct leading term for the exact result. We thus argue that pair production has its origin in the transition amplitude in the complex plane of time along a contour of winding number one.

\noindent {\bf Acknowledgments}\\
The author would like to thank Christian Schubert for useful discussions on particle production.
This work was supported by Basic Science Research Program through the National Research Foundation of Korea (NRF) funded by the Ministry of Education (NRF-2012R1A1B3002852).

\end{document}